\documentclass{iopart}

\newcommand{\be}{\begin{eqnarray}}
\newcommand{\ee}{\end{eqnarray}}

\newcommand{\beq}{\begin{equation}}
\newcommand{\eeq}{\end{equation}}
\newcommand{\ket}[1]{\left| {#1} \right>}
\newcommand{\bra}[1]{\left< {#1} \right|}

\usepackage{graphicx}

\begin{document}

\title{Context, spacetime loops, and the interpretation of quantum mechanics}

\author{Andrew M. Steane}
\address{
Centre for Quantum Computing,
Department of Atomic and Laser Physics, Clarendon Laboratory,\\
Parks Road, Oxford, OX1 3PU, England;\\
Universit\"at Ulm, Albert-Einstein-Allee 11, D-89069 Ulm, Germany.
}

\date{\today}

\begin{abstract}
Three postulates are discussed: first that well-defined properties
cannot be assigned to an isolated system, secondly that quantum
unitary evolution is atemporal, and thirdly that some physical
processes are never reversed. It is argued that these give useful
insight into quantum behaviour. The first postulate emphasizes the
fundamental role in physics of interactions and correlations, as opposed
to internal properties of systems. Statements about physical
interactions can only be framed in a context of further interactions.
This undermines the possibility of objectivity in physics. However,
quantum mechanics retains objectivity through the combination of the
second and third postulates. A rule is given for determining the circumstances
in which physical evolution is non-unitary. This rule appeals to the absence
of spacetime loops in the future evolution of a set of interacting systems.
A single universe undergoing non-unitary evolution is a viable interpretation.
\end{abstract}

\maketitle


\section{Introduction}

It is well established that the physical world is quantum mechanical. This
is established not only by carefully controlled experiments designed to
demonstrate basic phenomena such as interference of de Broglie waves,
and Bell-EPR correlations, but also
by the success of theoretical developments, such as Dirac's deduction of the existence
of anti-matter, and more complex insights such as quantum field theory and the
predicted existence of further particles such as the $Z$ boson, which were
subsequently detected experimentally. Despite this, there remains a fundamental
difficulty: there is no consensus on the clearest way to set out the
basic physical content of the theory of quantum mechanics. That is to say,
we understand how to use the theory for all practical purposes, but this is
done by making free use of loosely defined words such as ``measurement''.
If pressed to state exactly what physical process constitutes a measurement,
physicists experience varying degrees of satisfaction with their own answer,
but no-one's answer has commanded the sort of near-universal agreement which
we normally expect in science. This proves, in my opinion, that we have not
yet understood this subject properly.

I believe it is likely that a thorough resolution of this difficulty will only be possible
once a more general theory has been developed, such as one unifying
quantum mechanics and general relativity. It may also be that the whole
reductionist approach, though a useful method for simple systems, is
limited in scope, and not capable of treating some phenomena in sufficiently
complex systems. Even so, we should seek a reductionist description if one is available,
on the principle of not introducing unnecessary hypotheses.

In this article I will set out a symmetry principle which is not
commonly taught or emphasized in physics but which, I will argue,
should be given a more prominent position. It is obeyed by quantum
mechanics and not by classical physics, and gives a useful insight
into the former. I will also comment on reversible behaviour. I
will argue from the combination of these ideas for a specific
physical interpretation of quantum theory. That is, I will set out
a way to link the abstract mathematical apparatus of the theory to
statements about physical phenomena. In particular, I provide a
rule for determining the circumstances under which a `quantum
event' occurs, where a quantum event is a non-unitary evolution
roughly equivalent to a `collapse of the wavefunction'. The
discussion is like the Copenhagen Interpretation (CI) of quantum
mechanics \cite{28Bohr,BkPeres,BkShankar}, but seeks to avoid
unsatisfactory elements of the latter, especially its use of
concepts such as ``measurement'' or ``classical apparatus''
without a satisfactory definition. My approach also has an element
reminiscent of the ``transactional interpretation'' of J. Cramer
\cite{86Cramer,05Cramer,06Marchildon}, namely the idea that some
aspects of physical behaviour are atemporal, and correlations are
established by a combination of local interactions and a specific
type of influence from the future.

The symmetry principle, which I call the principle of `contextuality', is the
assertion that physical entities cannot have physical
properties in and of themselves. Interactions and correlations between entities are
more fundamental, and properties such as mass, velocity, etc. arise by a type
of symmetry-breaking. Basic theories of kinematics and dynamics must respect this
symmetry and this may be regarded as a partial explanation for some of the basic
features of quantum mechanics.
Correlations can be regarded as invariants of the associated transformation.

The idea that relationships and correlations are the only proper subjects of physics
is suggested by Everett's ``Relative State" formulation of quantum mechanics
\cite{57Everett}, it was briefly discussed by Zurek \cite{82Zurek}
and has been stressed more recently by Rovelli \cite{96Rovelli},
Mermin \cite{98Mermin} and others \cite{BkCushing}; see also \cite{BkdEspagnat}.
However there is a further ingredient to experimentally
observed behaviour that is not satisfactorily treated in these works, in my opinion.
This is the everyday observation that the Universe evolves in a non-unitary way.
Quantum theory correctly predicts the type and degree of correlation to be expected between
systems, but physical systems express these correlations by randomly adopting
physical configurations drawn from an appropriate set: for example, alive or dead,
in the case of Schr\"odinger's cat \cite{35Schrodinger}.
Therefore although I agree the ``relational'' symmetry is important
to understand the quantum formalism, I claim it is broken in the actual dynamics.
I introduce the word ``contextual'' because I will be concerned with dynamics
involving groups of three systems, not two.

In the following
I aim to elucidate this symmetry-breaking behaviour, not by extending quantum theory,
but by providing a rule for determining the circumstances under which
it occurs. The rule involves an appeal to processes that are not reversed,
c.f. \cite{BkHeisenberg,BkvonNeumann}, but does not
assume such processes are not treatable
by standard quantum theory. Instead, it is asserted that unitary evolution
is essentially atemporal, and quantum processes are sensitive to the absence
of loops (to be defined) in their future. One can then consistently claim
that physical
evolution consists of a sequence of non-unitary transformations or
`quantum events', with probabilities given by the standard formalism, and
preferred basis related to the idea of contextuality.

Although I will draw links between this `quantum event postulate'
and the other ideas presented here, it is nevertheless a
mere assertion, and this may be a weakness of the present account. A mechanism would
greatly clarify matters, if one exists, such as the transactional idea \cite{86Cramer},
the Ghirardi-Rimini-Weber or other stochastic mechanism \cite{86Ghirardi,03Bassi,97Fivel} or a
gravitational effect \cite{00Penrose,BkPenrose}. The present discussion leans towards the former
(transactional) account, which is an interpretation not an extension of quantum
theory, without committing itself to any particular view. That is, the approach
presented here does not require any fundamental new dynamics, but it can be
formulated so as to be consistent with some types of new dynamics
(see section \ref{s:false}). Generalisations
of quantum dynamics, such as non-linear terms in Schr\"odinger's equation, can
have the result of implying that some physical processes are computationally very
efficient, for example collapsing the complexity class NP to P. This would be
very surprising and when it happens it may merely indicate that the attempted
generalisation is wrong. Here I take the view that progress in formulating
dynamical equations will require mathematical tools able
to handle a dynamical spacetime (i.e. quantum gravity), but interpretive
discussions such as the present one can help in placing constraints on
and identifying desirable features of such a theory.

I will present the postulate on quantum events in a form where no new dynamics
are assumed. It has been argued
by Marchildon that Cramer's transactional interpretation makes correct predictions,
but retains one of the problems of CI,
namely the difficulty of making a ``quantum--classical" type of
distinction \cite{06Marchildon}. 
The quantum event postulate provides the required distinction.
Another approach to quantum mechanics which does not change the dynamical equations
is the ``consistent histories" or ``decoherent histories'' formalism of Griffiths, Omn\`es,
Gell-Mann and Hartle, see for example \cite{96Dowker,04Halliwell,06Hartle} and references therein.
Brief comments on the relationship between those ideas and the ones presented
here are given in section \ref{s:states}. The 
mathematical notion of a set of ``consistent histories"
appears to be able to describe a wide range of physical behaviour, and
has to be carefully interpreted. The degree to which interpretive statements
need to be added to the formalism is a matter of continuing debate \cite{97Kent,98Kent}. For example,
it is argued in \cite{96Dowker,95Dowker,98Kent} that there exist a large number
of consistent sets of histories, only a small proportion of which correspond
to the type of quasi-classical behaviour we observe in the world. Therefore a further
set selection criterion, beyond consistency, is needed. The ideas presented
here do not require a ``consistent histories'' formulation, but in that
formulation they would provide a further set selection criterion.

Sections \ref{s:behave} and \ref{s:rel} introduce the general flavour
of the ideas. In the former I sketch
the overall pattern of behaviour observed in experiments
sensitive to quantum phenomena, and comment on different mathematical
approaches to time evolution. The observations about reversible
evolution open the way to an appeal to atemporal behaviour which is
used in the arguments later in the paper.

Section \ref{s:rel} discusses the Principle of Relativity, in order to use it
to illustrate the role of a symmetry principle in
furthering our understanding of a physical theory. It introduces
the general idea of the need for a certain subtle type of economy in physical
theories, namely that a basic theory should not imply, even indirectly, that
physical entities can have unmeaningful characteristics, such as absolute velocity.
Section \ref{s:isolate} then introduces the general idea that it is questionable
whether properties ought to be attributed to isolated physical entities, and
it illustrates how this is handled in quantum theory.
Section \ref{s:context} introduces the three postulates which underpin the
main argument. Section \ref{s:quant} is the main part of the paper, it
applies these ideas to the interpretation of quantum mechanics, and introduces
the rule for quantum events. Section \ref{s:trans} presents a rough sketch
of a way of thinking about the rule, in the spirit of the `transactional'
mechanism.
A brief comment on the 2nd Law of Thermodynamics is made in section \ref{s:thermo},
and then section \ref{s:false} discusses the issue of possible falsification
of these ideas, and \ref{s:conc} concludes.

\section{Quantum mechanical behaviour}  \label{s:behave}

Experimental observations, such as the Young's slits experiment
with a low-flux source, invite us to relinquish the classical vision of a
spacetime which simply exists and has the worldlines of all the particles of
the universe in it. Instead, physical reality consists of a richer
behaviour which is harder to describe, because things really
`happen', in the following sense. In a reversible evolution, time
is merely a parameter, and whether it goes forward or backwards is of
no fundamental significance: the future causes the past in just
the same way as the past causes the future, in the sense that a
given situation at a time $t_f$ defines what the situation must
have been at earlier times $t_i$. The classical vision of
worldlines laid out in spacetime in what amounts to a sort
of `permanent present' is consistent with this. However, Nature
is not like that. There is genuine freedom for novelty in the
universe. Some events which might have happened do not happen, (e.g. the death
of a cat) and others which might not have happened, do (e.g. the cat survived).
The initial and final situations do not uniquely prescribe one another.
The universe lurches step-wise into the future like a wobbly child
picking her way across a stream on stepping stones.

In my opinion, part of the difficulty of understanding these things
lies with the fact that the language of state vectors and Schr\"odinger's
equation, while very useful for formulating calculations, is frankly
misleading when it comes to getting a good physical sense of what is
going on. It has two problems. First it focusses too strongly on the idea of a
`state' of a system, whereas we need to make interactions between systems
the central idea. Secondly, the notion of gradual evolution through time,
described by a differential equation, while it can be pressed into
use with sufficient interpretive statements accompanying it, is not the
natural language to describe the real world.

To illustrate the second point, consider classical mechanics.
This can be mathematically treated both by a differential
equation describing gradual evolution through time, and
also by a path integral (least action)
method. The predictions are the same and in the classical case
there is no difficulty in
making a close comparison between the two physical pictures suggested by
the mathematical equations. However, in the quantum case, I will argue,
the path integral method gives a better physical picture than the
time-dependent differential equation method. This is
because it forces one to consider a physical problem from the point
of view of final as well as initial conditions, and it presents a global
view of the worldline. The formulation presents the final condition
as a {\em fait accompli}, and asks for its probability.
The global view of the worldline
emphasizes that time acts simply as a parameter during reversible
evolution, and there is no way to pick its direction without further
information.

In the following, I will regard the paths entering into a path integral
as (highly structured) links between different points in spacetime,
but physical evolution is non-unitary. Processes in the present are
sensitive to the absence of a loop (of a type to be defined)
in their future
light cones. We avoid causal-loop paradoxes or contradictions
by ensuring that the physical
predictions are those of standard quantum theory.

\section{The Principle of Relativity} \label{s:rel}

It is useful to clarify what {\em type} of explanation is to be
put forward here. I will not be proposing any new equation, but
rather offering a new perspective on familiar phenomena.

A useful comparison can be made with the type of explanation
provided by Einstein's special theory of relativity \cite{05Einstein}.
At the time
Einstein put forward his ideas, it was already known that objects
in motion relative to a given reference frame should be contracted
and evolve more slowly, etc., which is why the Lorentz contraction
and the Lorentz transformation have been named after Lorentz
rather than Einstein. Indeed, if one assumes that Maxwell's
electromagnetic equations are correct, then the special theory of
relativity makes no new predictions for electromagnetic phenomena
(which include the dimensions of everyday objects, the mechanism
of everyday clocks, and so on). However, Einstein's theory
provides a profound and very useful insight,
because it shows that the Lorentz transformation arises in a
simple and general way from a few reasonable assumptions about
Nature. This both clarifies what the Lorentz transformation
means (i.e. it is a statement about
space and time and not about electromagnetism {\em per se}), and
it allows one to expect and require not just
electromagnetic theory but any theory of the natural world to be
Lorentz covariant.

Einstein singled out two principles: relativity, and the principle that
the speed of light in vacuum is independent of the motion of the source
\footnote{One does not have to refer to light, it is
sufficient to claim that there is a finite maximum speed for signals.}.
He then insisted that these two, which appear at first to be
contradictory, are in fact mutually consistent, as long as one lets
go of the mistaken idea that simultaneity is absolute.

Any interpretation of quantum theory which does not involve an extension
to the formalism must make an argument of this broad type, in that the
difficulty is in reconciling things which appear to be contradictory,
and to do this one must let go of some mistaken preconception. The
apparently contradictory things in quantum mechanics
are the unitary equations of motion
and the observed non-unitary physical behaviour. In a `many-worlds'
type of interpretation \cite{BkDeWitt}, one tries to reconcile these by
letting go of the notion of a single universe, or of an observer-independent
set of physical events. However, it is not clear that this succeeds,
because one needs irreversible behaviour to cause a `split',
which is begging the question ({\em petitio principii}). Rovelli's
`relational' interpretation has some similarities
with `many-worlds', and suffers, in my opinion, from a similar problem.
In \cite{96Rovelli} the preliminary discussion appeals to `a specific
measurement outcome' and `the standard account of measurement'.
Admittedly the full discussion is more thorough, but it still has
to use terminology such as `a quantum event' involving
`discrete changes of the relative state,
when information is updated' \cite{06Smerlak}.

I will advocate that one should retain a
single universe with observer-independent physical
events, but allow a more subtle relationship between the present and
the future than is the case in classical physics.

The idea of contextuality, to be discussed below, is
similar to the relativity principle. To bring out the similarity
it is useful to state the Principle of Relativity in the form
 \nopagebreak
\begin{quote}
{\bf Principle of Relativity of Kinematics}: {\em The Laws of
Nature should take a form such that only relative uniform motion, not
absolute uniform motion, can be meaningfully defined.}
\end{quote}
The related statement,
``the Laws of Nature must take the same mathematical form in all inertial reference
frames," can be regarded as following from the former.
I have called this the ``Principle of Relativity of Kinematics''
rather than simply the ``Principle of Relativity'' since I will be
discussing a broader type of relativity principle below.

The Principle of Relativity of Kinematics
takes a logically satisfying intuition about
the way physical systems can be expected to behave, and proposes it as a Law
of Nature. In particular, it claims that a statement intended to
describe or quantify uniform motion can only be framed in
terms of the relative motion
of one body with respect to another. From this it follows that the
mathematical expression of the laws
of physics should not imply that a further, absolute, type of uniform
motion is detectable, or has physical consequences of any sort.
I propose to take this idea further, and claim that not just
uniform motion, but every aspect of physical reality can only be defined
in a relative way, in the following sense. First, there are no
properties of any physical entity in and of itself;
rather, `properties' are a useful way to summarize collections of
interactions between entities. These interactions, and the
correlations between entities which they produce, are the
fundamental elements of physical reality. Furthermore, these
`fundamental elements' (i.e. the interactions and/or correlations)
cannot be defined in an absolute way, but rather they can be
specified only relative to other sets of interactions and
correlations, which I will call their {\em context}.
However, some aspects of the context are permanent, and this allows
objective events to occur. I will
clarify the meaning of these statements after I have given some
further argument in support of this general approach.

\subsection{Isolated systems are ill-defined}  \label{s:isolate}

Suppose there only exists, in the whole of reality, one simple
particle. What statements can be made about such a particle?
None (except the assumed fact of its existence). Obviously
there is no way to define its location and speed, since the
notion of absolute motion is ill-defined (i.e. meaningless),
and there is nothing with respect to which it can have a relative
motion or position. Similarly, only relative mass and charge and
so on is meaningful.

It may be objected that if the whole of physical
reality were to consist of a single simple particle, one
could not reason about it in any case.
However, in the real universe there can exist an
approximation to the above situation, namely a simple particle which
is isolated from all other things for a long time. To the extent
to which such a particle is in fact isolated, and remains so,
we should therefore expect the Laws of Nature to describe it in
such a way that all its properties are undefined. Classical
physical theories could not do this, because their starting point
is the concept of entities with well-defined properties. However,
quantum mechanics does offer a mathematically consistent way
to handle such a possibility.

The way in which the properties of an isolated particle are not
well-defined in quantum mechanics
is not merely a matter of quantum uncertainty, a spread in
the wavefunction. It is that one cannot assign a quantum state to
such a particle. When we make statements such as ``a particle
is prepared in the state $\ket{\phi}$" what we mean is that the
particle undergoes an interaction with another system (typically
large and complicated such as an absorbing barrier)
such that the particle and the other system are entangled, and
the evolution to be described concerns only one part of this
entangled state. However, if the part of the entangled state
which was ignored is later caused to interfere with the part
which was under discussion, then the whole discussion was invalid
because the premise (``particle prepared in $\ket{\phi}$") is false.
For an isolated particle, there is always
the possibility that it will in future couple to
systems with which it is currently entangled, so that it is
impossible to make well-defined statements about its
quantum state without reference to such other systems.

A simple example of a physical property that is
undefined, and indeed meaningless,
is the spin state of a spin-half particle that is
one of a pair of particles in a maximally entangled state
such as the singlet:
\be
\left( \ket{\uparrow} \otimes \ket{\downarrow}
- \ket{\downarrow} \otimes \ket{\uparrow} \right)/\sqrt{2}.
\label{singlet}
\ee
Einstein, Podolsky and Rosen gave an argument to suggest
that an ``element of reality'' 
should be associated with the spin state of either particle individually,
but the fact that such a quantum system can give
rise to correlations which do not satisfy the Bell inequality
shows either that this assumption is false, or that each
spin ``element of reality"
is sensitive to distant apparatus settings \cite{35Einstein,BkBell,BkPeres}.
I maintain the former, i.e. the term ``spin state" simply cannot
be applied to each individual particle in a singlet (except to
say it is part of the singlet).
Locality and the concept of a ``state" are discussed further
in section \ref{s:states}.

The analysis offered by quantum mechanics, which we write
down using a mathematical notation such as (\ref{singlet}),
offers a precise way to express the notion that, in appropriately
prepared circumstances, the assignment of a value of a property, in this
example the direction of spin angular momentum, to an individual
entity (one of the particles), can be meaningless, even though
there are other circumstances where such an assignment can be made.

The notation in the Schr\"odinger picture, eq. (\ref{singlet}),
is unfortunate in that it forces us to write what might appear to be
individual states for the two particles, suggesting to some
that the spin of either particle is `partly up' and `partly down'. This
is merely a limitation
of notation, however, or a case of over-interpreting mathematical symbols.
One must simply refrain from trying to speak as if properties can be assigned
to individual systems\footnote{We will discuss later the circumstances under
which such language can be allowed because the context makes it unambiguous.}.
The fact that the spin is not `partly up' and `partly down' is underlined by
the fact that there is no way one can legitimately
choose the `up' and `down' directions
instead of some other pair of directions, because of the well-known rotational
symmetry of the singlet state, e.g.
\[
\frac{1}{\sqrt{2}}(\ket{\downarrow \uparrow} - \ket{\uparrow
\downarrow})= \frac{1}{\sqrt{2}}(\ket{\rightarrow \leftarrow} -
\ket{\leftarrow \rightarrow})
\]
where $\ket{\leftarrow} = (\ket{\downarrow} + \ket{\uparrow})/\sqrt{2}$,
$\ket{\rightarrow} = (\ket{\downarrow} -
\ket{\uparrow})/\sqrt{2}$, and I use the shorthand
$\ket{m_1 m_2} \equiv \ket{m_1} \otimes \ket{m_2}$
. In the language of quantum mechanical interpretation
discussions, we say there is no `preferred basis'.

Another instructive way to present this rotational symmetry is as follows.
Suppose we first rotate just one of the particles. Then in order to restore the ket
to its initial form, one can rotate either of the particles: either reverse
the rotation of the 1st particle, or apply the same rotation to the 2nd. Therefore,
a given transformation of the composite system
can be accomplished by rotating either one of the constituent particles.
A similar symmetry applies to all the Bell states \cite{05Zurek}.
It is obvious that such behaviour is not possible for a pair of classical arrows,
and it implies that it cannot be correct to discuss the composite system
as if the two constituent particles contributed individual spin properties.
I am labouring this point because it is particularly striking when one
recalls that the two particles can be space-like separated.
It may also bear on the computational power of quantum computers: I have
argued elsewhere that the advantage available to quantum
computing, compared with classical computing, arises
from exploiting precisely this feature, namely that quantum systems can
express and manipulate a physical representation of the
correlations between logical entities, without the need to completely
represent the logical entities themselves \cite{03Steane2}.

The feature of quantum mechanics which allows it to provide this
type of description is, of course, entanglement.
Entanglement is the means by which the Laws of Nature are consistent with
the requirement that mutual influences and correlations are more
fundamental than properties of isolated entities.

Why is it the case, then, that assigning properties to entities
(``the cup is blue, the ball is heavy," etc.) is so
thoroughly built in to almost all our reasoning about the physical world?
This is because such statements are made within a context, namely the
actual history and future of the world, and properties are emergent phenomena.
That is, physical behaviour tends towards a situation where associating
specific properties with individual entities is valid.
I will clarify this in the next sections.

\section{Contextuality, atemporal evolution and irreversible processes}  \label{s:context}

I will now state the physical principles that I wish to put forward,
and whose implications are the main subject of this paper.

\begin{quote}
{\bf Postulate 1} (``Contextuality"): {\em The Laws of Nature should
take a form such that well-defined properties cannot be
assigned to an isolated system. Only interactions between
systems are meaningful, and these can only be described through
their influence on subsequent interactions of the systems
in question with the rest of the world.}
\end{quote}

\begin{quote}
{\bf Postulate 2} (``Atemporal evolution"):
{\em Quantum unitary evolution is atemporal.}
\end{quote}

\begin{quote}
{\bf Postulate 3}  (``Irreversibility"): {\em There are in Nature processes
that are not reversed.}
\end{quote}

To many physicists, the first and third of these assertions may be
unremarkable, even obvious.
However, it is remarkable that, in conjunction with
the basic equation of motion (furnishing the propagator for unitary evolution,
for example by a path integral) they suffice to allow a physical
interpretation of quantum theory.
What is meant by the second postulate, on atemporal evolution, will be
explained in the following. I have included it in this list in order
to highlight it, and to bring out the tension between this postulate and the
third (irreversibility).

The irreversibility postulate refers to ``processes that are not reversed''.
The discussion will not involve
an irreversible component to the fundamental equations of motion, such as
in dynamical wavefunction collapse theories \cite{BkPenrose,03Bassi,86Ghirardi,00Penrose,97Fivel}.
For a process to be called ``non-reversed'' here, it is sufficient if
the equations of motion are
reversible but the motion is such that it never gets reversed
in practice. For example, it could involve a particle emitted
to very large distances, or it could be very complex.

Even supposing that the equations of motion are
reversible, it is clear that many physical processes are in
actual practice not reversed on any time scale to which we are able
to assign meaning (e.g. the lifetime of galaxies, the proton decay time).
Furthermore, it might be strictly impossible to reverse a very complicated
process such as an avalanche, because any
apparatus intended to reverse the motion of all the
particles would itself be large and complicated,
and could not be sufficiently isolated. For example, its
gravitational influence on the rest of the world would be especially
hard to avoid or reverse.

Once we insist on a context to statements about physical behaviour,
there immediately arises the possibility of ambiguity.
For example, relative motion is a well-defined concept because
the relative motion between one body and another can be specified
without the need to bring in a possible relative motion of the second
body with respect to a third. However, if a statement about relative
motion is only meaningful within a context, then
a third body might be important after all, because the context
could depend on it.
This is exactly what is investigated in well-known
paradoxical experiments that have been long discussed in
quantum mechanics, such as ``delayed-choice'' experiments, the
``quantum eraser" and ``Wigner's friend" \cite{BkPeres,BkWheeler,86Cramer}. We will
avoid ambiguity about physical behaviour by using postulates 2 and 3.
This will be discussed in section \ref{s:quant} below.
I will also argue that these three postulates are intimately linked.

By requiring that physical systems cannot be regarded as isolated entities,
the contextuality postulate
places a constraint on the form of other Laws of Nature. In common
with what is found in general when physical behaviour is subject
to a constraint at a fundamental level, we may guess that
physical behaviour will be found in practice to fill the
constraint, i.e. satisfy it but only just. Therefore I predict
that physical systems will show a tendency to maximise the degree
to which they can be regarded as separate entities with individual
properties. This means they will minimise entanglement.  
This prediction is not unavoidably implied by postulate 1, 
however, therefore I will
propose it as part of a further postulate below (section \ref{s:basis}). 
The point is that the
further ``quantum event" postulate is not altogether independent of
1--3, but rather is suggested by them.

\section{Application to quantum theory} \label{s:quant}

\subsection{No properties for isolated systems}

Standard quantum mechanics obeys the contextuality postulate at
the most basic level of single isolated entities because
quantum entanglement implies that no
statements can be made about the properties of completely isolated
systems.

For example, consider the projection
onto some chosen direction (taken as the $z$ axis) of the spin
{\boldmath $\sigma$} of a given electron. We know that the
possible eigenvalues of the spin component $\sigma_z$ are $\pm
\hbar/2$.
However an electron, in and of itself, cannot be said
to have a value of $\sigma_z$ since its spin state might always be
entangled with something---possibly the last atom it scattered
off, or else if it never scattered off anything then a distant
positron (if the electron came from pair production), 
or a proton and anti-neutrino (if it
was a product of beta decay), or some quantum fields (if it
came from the Big Bang). We will discuss in the next section
the circumstances in which a `quantum event' can result in
a well-defined $\sigma_z$, but such an event depends on the interaction
of the electron with other things.

It is reasonable to assume that all properties, including mass and charge
and the total spin of particles, are generated by quantum effects,
in which case the argument applies to these properties also. When we
say an electron has a well-defined mass, which we regard as an intrinsic
property, this is because it has acquired a well-defined mass by virtue
of past events, involving interactions with other systems.
Different mass states were entangled with decay products in
the early history of the universe (just as an atom undergoing spontaneous
emission is entangled with the emitted photon), when evolution between
mass states of the fundamental entities (strings or whatever) took place.
However these decays are not going to be reversed in the future, and in
this circumstance (see next section) one of the mass states was
adopted randomly. One of the mass
eigenvalues is $\sim 9.10939 \times 10^{-31}$ kg, and entities
of this mass (and various other properties which come about in
analogous ways) we call electrons. We can talk about ``properties"
because we have implicitly assumed this history.

\subsection{Quantum events}   \label{s:basis}

The interpretation problem, or measurement problem, in quantum mechanics
is essentially the problem of wavefunction collapse. It is illustrated
by the Schr\"odinger cat experiment:
we want to know whether and how the
`both and' character of a superposition can be resolved into the
`either or' character of a set of possible outcomes.
CI handles
this by a statement that the whole theoretical formalism is there to
describe possible behaviours of `classical' systems, but it fails to
explain how these classical systems are identified, or why they are not
quantum systems. In the approach taken by Feynman in his famous lectures
on physics \cite{BkFeynmanLectures,48Feynman}, the problem is there but hidden. In
this approach, one identifies the final situation whose probability is
desired to be calculated (just as in a path integral calculation), and one goes ahead
and calculates it. It is not discussed how the physical system `knows'
to adopt just one of the possible outcomes and not all of them in superposition.
The discussion of Cramer \cite{86Cramer} makes the same omission. Cramer
appeals to the notion of a `quantum event' and discusses how it may
insightfully be understood as an atemporal `transaction' between emitter
and absorber. However, he omits to say how one identifies when a `quantum event'
occurs, as opposed to a unitary evolution in which several absorbers become
entangled. This point has also been raised by Marchildon \cite{06Marchildon}.

I agree with Cramer that a `quantum event' is best understood to
take place over an extended region in spacetime, not at a spacetime point.
This idea is implied in CI, but not clearly spelled out. The minimum formal apparatus
we need in order to interpret the theory is a statement to identify when
a `quantum event' occurs. I will now provide such a statement. The rest
of the paper is a discussion of its meaning and application.

\begin{quote}
{\bf Quantum Event Postulate} (``No-loop") \\
Strong version: {\em
A ``quantum event'' or ``transaction'' is undertaken whenever
there is no loop in the future whereby the relative phase of two
parts of an entanglement could influence further events.
The preferred basis is that in which separability is maximised.}\\
\\
Weak version:
{\em
A ``quantum event'' or ``transaction'' is undertaken whenever
there is no loop smaller than $A_{\rm max}$
in the future whereby the relative phase of two
parts of an entanglement could influence further events.
The preferred basis is that in which separability is maximised.}
\end{quote}

In this postulate, the term ``quantum event'' refers to a non-unitary
evolution from the present to the future, in which one of a set
of possible outcomes is realised. It corresponds roughly to the notion of
``measurement'' in CI.
The preferred basis defines from which orthonormal set the outcome is to be drawn
(randomly with probability equal to the modulus square of
the quantum amplitude). 
The absence of a loop recalls the well-known idea 
of (the presence of) ``which path'' ({\em welcher weg}) information, also
known as a {\em record}.
Under purely unitary dynamics, a system $W$ carrying {\em welcher weg} 
information would be entangled with the system $S$ whose path information 
it carries. The postulate applies when $S$ is itself composite,
and asserts, essentially, that if the future dynamics does not erase
the {\em welcher weg} information, and the latter can discriminate
between separable and inseparable states of $S$ (see below), then an 
event occurs, i.e. a 
non-unitary evolution is completed in a finite region of spacetime.
Separability here refers to
the absence of non-local correlations, i.e. it is a property of the physical
behaviour in space-time, not the abstract analysis of vectors in Hilbert space.
The spacetime area $A_{\rm max}$ is a non-trivial quantity whose
definition will be discussed in the following, as will the identification
of the preferred basis. 

I have employed
Cramer's term ``transaction'' in order to make it clear that the
``quantum event'' is extended over spacetime. The postulate is
independent of whether or not a microscopic machinery of ``offer wave''
and ``confirmation wave'' is assumed, but I sketch in section
\ref{s:trans} how such a machinery might be constructed.
The statement about preferred basis advances the hypothesis that
systems in practice tend to become separable, within the
constraints set by the equations of motion and the boundary conditions.

I will discuss the strong version of the postulate, and comment on
the weak version afterwards. I will use 2-state systems to illustrate
the physics, and refer to them as `spin-half particles'. The two
states in question don't have to be spin states, they could for
example refer to left and right motion, or ground and excited states
of some system. However the 2-state systems in question are small and simple.
I will use standard quantum mechanics in the Schr\"odinger picture to
treat the evolution mathematically, and I will show how the quantum event
(no-loop) postulate is suggested by or connected to postulates 1--3. The
discussion will treat a `toy' or simplest possible case, followed by
some comments on the extension to more general cases.

Consider two spin-half particles $A$ and $B$ which interact with one another.
Referring to `spin-half particles' is logically consistent because the total
spin of the physical entities under discussion, and some other
basic characteristics, will be well-defined by past processes.
In order to keep the problem simple, it will be assumed that the
particles are not identical, and their
motional states are small wavepackets which can
be approximately treated as classical particles moving along
classical trajectories. This is consistent when the
interactions of the particles with the rest of the universe have
resulted in well-defined motional states, and the further evolution
under discussion does not
entangle their motional degrees of freedom.

In view of the postulate (contextuality)
that physical statements must not imply that isolated particles
have absolutely defined properties, one must be careful
not to use the word `state' inappropriately. Therefore when
referring to the mathematical apparatus of vectors
in Hilbert space, I will use the word
`ket' rather than `state vector' or `state'.
Also the phrase ``in the context $\cal R$'' will be used as a shorthand
for the phrase ``in the context of interactions outside spacetime region
$\cal R$''.

Suppose that, in some spacetime region $\cal R$, the spins of the
particles are initially (i.e. where the world lines enter $\cal R$)
described by the ket $\ket{\leftarrow} \otimes \ket{\downarrow}$
(we will examine at the end how this can come about).
Suppose first that they evolve under an interaction between them,
such as $\sigma_x \otimes \sigma_x$,
but they do not interact with anything else.
In this case, by postulate (contextuality),
there are no physical predictions to make. This is because we need at
least three entities: two to have an interaction, and a third to be influenced
by the result. This places the interactions at a more basic level than
the entities interacting\footnote{The related idea in Cramer's analysis is
that a fundamental irreducible event involves the emitter,
the absorber, and the field. For example, if a 2-level atom emitted a photon,
then the transactional interpretation only allows a discussion of the outcome
when another atom is available to absorb the photon.}.

\begin{figure}[ht!]
\centerline{\resizebox{!}{0.25\textwidth}{\includegraphics{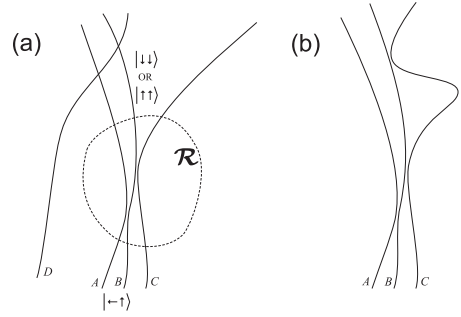}}}
\caption{Spacetime diagrams. (a) Particles $A$, $B$, $C$ interact and become
entangled. $A$ and $B$ subsequently interact
with other systems $D$, but not with $C$ or any system
influenced by $C$ (but see section \ref{s:discuss}). The resulting
evolution is non-unitary, placing $A$ and $B$ in one of $\ket{\downarrow\downarrow}$
or $\ket{\uparrow\uparrow}$ (before their further possible interactions with $D$).
(b)
Particles $A$, $B$, $C$ interact and become entangled, and
then $C$ subsequently becomes disentangled with the others.
This evolution is unitary and part of some larger event.
}
\end{figure}

Next consider the case that $A$ and $B$ begin in
$\ket{\leftarrow} \otimes \ket{\downarrow}$
and subsequently interact with each other and a third particle $C$
so as to evolve into a tri-partite fully entangled ket such as
$\ket{\psi_3(\phi)} \equiv (\ket{\downarrow \downarrow \downarrow}
+ e^{i\phi} \ket{\uparrow \uparrow
\uparrow})/\sqrt{2}$. We consider two cases for the further evolution
of the particles. In case (a),
$A$ and $B$ do not ever interact further
with $C$, or with any system influenced by $C$
(figure 1a). In case (b), $A$ and $B$ interact again with $C$ so as to
disentangle the latter, to produce the final ket
$\ket{\psi_2(\varphi)} \equiv (\ket{\downarrow \downarrow}
+ e^{i\varphi} \ket{\uparrow \uparrow}) \otimes \ket{\uparrow}/\sqrt{2}$ (figure 1b).

In case (b), the phase $\varphi$ is observable. That is, it can influence
events in the future. In case (a), the phase $\phi$ is unobservable
because in order for it to influence further events, an interference
or a correlation must
be brought about, but, by construction, this does not happen, because
we said $A$ and $B$ do not again interact with $C$ or any system influenced
by $C$.

By postulate (no-loop), in case (a) a quantum event takes
place, whereas in case (b) one does not---instead, the unitary evolution is
part of some larger event.

The preferred basis is identified as follows.
According to the contextuality postulate, we
want to discuss the interaction between $A$ and $B$ in terms of its
influence on future interactions. In view of the fact that $A$ and $B$ are not
further influenced by $C$, this can be done without reference to
$C$. Therefore we can get the complete information we need
after using a mathematical device to discard all
information associated with $C$. This device is the standard
(and arguably unique \cite{BkNielsen}) procedure
of averaging over the possible influences $C$ might have on some
further system. In mathematical terms, it
is the partial trace \cite{BkPeres,BkNielsen} of the density matrix,
$\rho^{AB} = {\rm Tr}_C[ \rho^{ABC} ]$. After this
averaging, we obtain for the ket of $A$ and $B$ a probabilistic statement
which can be expressed by means of the density matrix
  \beq
\rho^{AB} = \left( \ket{\downarrow \downarrow} \bra{\downarrow
\downarrow} + \ket{\uparrow \uparrow} \bra{\uparrow \uparrow}
\right)/2.      \label{sep}
\eeq
This density matrix can be written in terms of any set of basis states. For example,
expressed in the basis $\{ \ket{\rightarrow},\;\ket{\leftarrow} \}$ it is
  \be
\rho^{AB} &=& \frac{1}{4} \left( ( \ket{\rightarrow \rightarrow} + \ket{\leftarrow \leftarrow} )
                   ( \bra{\rightarrow \rightarrow} + \bra{\leftarrow \leftarrow} ) \right.
  \left.    +  ( \ket{\rightarrow \leftarrow} + \ket{\leftarrow \rightarrow} )
                   ( \bra{\rightarrow \leftarrow} + \bra{\leftarrow \rightarrow} ) \right) .
\label{ent}
  \ee
It is well established that the density matrix contains all the information
about the relevant systems ($A$ and $B$) needed to discuss their future in terms
of observables and expectation values, as long as they are not
entangled with something else. It does not,
in and of itself, define a preferred basis: it can be decomposed
(expressed as a sum of pure density matrices) in infinitely many ways.
However, the property of separability can distinguish one basis from
another, for a density matrix of a composite system. By postulate,
we now pick the basis in which the terms in the decomposition are separable,
i.e. in this case $\{ \ket{\uparrow\uparrow}, \ket{\downarrow\downarrow} \}$
rather than the Bell basis. The quantum event evolves $A$ and $B$ to
one of these final states, where now the word `state' is well-defined,
it refers to the initial condition of the next quantum event
involving the spins of these particles.

Although the term `quantum
event' suggests an abrupt process, I have already emphasized that it
is best understood to be extended in spacetime, and one can further
remark that it is not necessary to be precise about when it begins and ends.
It is sufficient to take the final condition of one event to be the
initial condition of the next.

There is an important distinction to note here, between different uses of the
density matrix. In a physical process involving entanglement whose presence
can be observed, such as case (b) above, one has an inseparable joint ket
describing two or more systems. In this case one can use the partial trace
as a mathematical tool to study the composite system, and indeed this
is commonly done in quantum information theory and is a useful tool.
Such a reduced density matrix is different from the one we used in case (a)
however, where it was a device to help identify the preferred basis when
a quantum event occurs. In eq. (\ref{sep}) the mixture represents the
`ignorance' of the universe, before the event, of which outcome will
occur, and this is also our ignorance, after the event but before we learn
the outcome. In case (a) the density matrix gives complete information on
what can be said about the outcome before the event. In case (b) any
reduced density matrix one may care to calculate (describing part of a
composite entangled system) gives incomplete information. This
distinction is not drawn in \cite{98Mermin}.

It is now possible to specify more precisely what was meant by our opening
statement that ``the spins of the particles are initially described by the
ket $\ket{\leftarrow} \otimes \ket{\downarrow}$". This means
that the spins had previously interacted with other systems such that the
result was a quantum event in
which this was one of the terms in the preferred basis (that which maximised
separability), and in the event the physical configuration
adopted was the one described by this ket. Furthermore, it was legitimate
to talk of `particles of spin half', because this is a shorthand for a
history in which the particles earlier (e.g. in the Big Bang)
interacted with other systems in such a way as to make their total
spin well-defined.

\subsection{Discussion} \label{s:discuss}

The toy example, figure 1a, considered a case where the 3rd particle, $C$,
never influenced $A$ and $B$ again in any way at all. This was to keep
the discussion as simple as possible. More generally, the same conclusions
would hold if any further influence (direct or indirect)
of $C$ on $A$ and $B$ is not such as to make the phase $\phi$ observable.
In practice, this could come about by a number of ways. The simplest
is where $C$ propagates to infinity, but this is arguably unphysical.
More usually, it would be because $C$ interacted with further particles
and initiated a non-reversed process such as an avalanche in the
$\{ \ket{\downarrow}, \ket{\uparrow} \}$ basis,
e.g. the propagator
$\ket{\uparrow}_C \ket{\Uparrow}_D \rightarrow
\ket{\uparrow}_C \ket{\Uparrow}_D$,
$\ket{\downarrow}_C \ket{\Uparrow}_D \rightarrow
\ket{\downarrow}_C \ket{\Downarrow}_D$, where $D$ is a large system
of many particles. The literature on the `environment-induced decoherence'
is a study of this type of process, see
for example \cite{04Schlosshauer} and references therein.
Decoherence, in which the off-diagonal elements of a density matrix
are zero although the related populations are not, does not on its
own solve the interpretation problem, but it allows one to discover
circumstances in which a relative phase is going to be unobservable in
the future because the required loop in spacetime involves a reversal
of complex behaviour. One then needs to assume that such a process
is indeed not going to be reversed (irreversibility postulate), and
then apply the quantum event (no-loop) postulate
(or make some other statement).

As an example of a slightly more complicated process in which the
phase $\phi$ in $\psi_3(\phi)$ is observable, consider the following.
First recall the identity
\be
\frac{ \ket{\downarrow \downarrow \downarrow} + e^{i\phi}
\ket{\uparrow \uparrow \uparrow } }
{\sqrt{2}} = \frac{
\ket{\phi_+}_{AB} \ket{\leftarrow}_C + \ket{\phi_-}_{AB} \ket{\rightarrow}_C } {\sqrt{2}}
\ee
where $\ket{\phi_\pm} =
(\ket{\downarrow \downarrow} \pm e^{i\phi} \ket{\uparrow \uparrow}) /\sqrt{2}$.
We introduce some further particles, and consider the following further evolution
of the joint system:
\be
\ket{\psi_3}_{ABC} \ket{\downarrow\downarrow\downarrow\downarrow}
&\longrightarrow&
\left(
\ket{\phi_+}_{AB} \ket{\leftarrow}_C \ket{\downarrow\downarrow\downarrow\downarrow}
+ \ket{\phi_-}_{AB} \ket{\rightarrow}_C \ket{\uparrow\uparrow\uparrow\uparrow} \right) / \sqrt{2}.
\label{mright}
\ee
In traditional language, (\ref{mright})
could be the start of a measurement of $C$ in the
$\{ \ket{\leftarrow}, \ket{\rightarrow} \}$ basis.
The eventual measurement outcome could be found to have perfect correlation with
the result of a measurement of $A,B$ in the Bell basis, leading to the conclusion
that $A,B$ were properly described by $\ket{\phi_+}$ or $\ket{\phi_-}$, not
$\ket{\uparrow\uparrow}$ or $\ket{\downarrow\downarrow}$, after
their interaction with $C$. Scenarios such as this
have been long discussed, and I don't want to rehearse that discussion here.
The central point is that all such observations involve a loop in spacetime:
in this example, the loop includes the `classical' transmission of the
measurement outcome from one place to another, in order to allow the correlation
to be revealed. Such a loop is precisely the one referred to in the
quantum event postulate. The quantum event, or, if you prefer, collapse
of the wavefunction, or transaction, occurs precisely in those circumstances
where no future process will probe the presence of an entanglement.

If there is a spacetime loop, then usually it is not the only possible
future: the unitary evolution includes other paths which do not form
a loop. This simply means that more than one type of quantum event is
available. Each possible event (i.e. each case having a no-loop future)
has a well-defined quantum amplitude, and
is picked with the corresponding probability. This agrees with the standard
predictions which one could arrive at, for example, via CI, but we have replaced
the notion of `measurement' by its underlying ingredients: non-erased
{\em welcher weg} information, and evolution towards separable states.

It is dangerous to use the word `never' in physics, but we have done so
(implicitly) twice: in the postulate on irreversibility and again in
the postulate on quantum events (the absence of loops in the future).
There are two ways to avoid an appeal to the infinite future here.
First, if it is possible to identify a non-reversed process in a finite time,
that would be sufficient for the strong version of
the event postulate. Secondly, one can imagine that
quantization of spacetime, such as in loop quantum gravity, might make
the interference phase undefined for finite but
very large or complicated loops in spacetime: this would be sufficient
for the weaker version of the event postulate. The loop area bound
$A_{\rm max}$ would be a complicated function of the behaviour of
all the systems involved in a large entanglement.

In practice it is easy to identify at least some
processes which we can be close to certain will never be reversed. The
traditional ``measurements'' such as absorption of a photon by a barrier
are among these.

This concludes the resolution of the measurement/interpretation
problem in quantum theory. In the Schr\"odinger cat
paradox, we conclude that the cat really is either alive or
dead, not both, and outcomes of ``measurements'' are
well-defined as either one outcome or another,
because they are associated with non-reversed
entanglement\footnote{There is no need for any conscious observer
in these discussions. The interpretation correctly predicts the outcome
of the Schr\"odinger cat experiment, {\em viz.} the objective reality
of a cat {\em either} alive {\em or} dead. If we
happen not to be aware of which eventuality has come about, then we can choose
to represent our best knowledge in terms of classical probabilities for the
outcomes. If we become aware of which eventuality has happened, then
of course the probabilities we assign must change, in the same way they
will change when a classical die is thrown under a cup, and we lift the cup.}.

One may say that the postulates 1--3 `work together' in the following sense.
The contextuality postulate on its own does not allow
an unambiguous interpretation of quantum theory, because it implies
that physical behaviour depends on context. One is left with a universe
apparently unable to have any objective physical behaviour.
It seems to require therefore the irreversibility postulate, as a minimal
statement that something objective can happen in the universe.
In order to satisfy both postulates, quantum theory involves a combination
of atemporal behaviour, where time is merely a parameter, and temporal
behaviour (the irreversible quantum events). The temporal behaviour
occurs whenever it is consistent with the topology of the
atemporal worldlines. The overall result is that
while the contextuality postulate constrains physical behaviour
so as to prevent assignment of properties to isolated systems,
systems behave in practice in
such a way that properties can be assigned to them as much as possible.

\subsubsection{Quantum states, locality} \label{s:states}
\noindent

We already showed that the concept of a ``state" is inappropriate to
degrees of freedom of a given system which are entangled with other systems.
We now discuss a more general limitation to the notion of a ``state" 
in quantum mechanics.

When we refer to a physical ``system" we are taking the step of referring, for purposes
of discussion, to some set of objects such as the particles in
some spatial region, following the principle that this is useful 
because of reductionism. In
referring to a ``state" of a system, we are making a similar notional separation,
but now with regard to time instead of space. By the ``state" of a system, we
mean generally whatever information is enough to specify the outcome of any
interaction with any other system, such that the only further information needed
is the state of the other system and the type of interaction. If two different preparation
processes bring two similar systems to a situation such that, if
they were subject to the same future interactions with third parties
then the same outcomes
would occur, then we say the two systems were prepared in the same state.
This is an important idea because the information needed to specify a state
can be finite, and need not involve the details of the past
history or future evolution of the system. In fact, the concept of a ``state"
in classical physics serves to identify just those aspects of a system which
can be assigned to a specific time.

In quantum physics it has been common to refer to the system ket as a ``state vector",
or simply a ``state'', on the grounds that it is sufficient to allow the
calculation of any experimental outcomes in which we may take an interest.
Thus, if a ket $\ket{\psi(t_i)}$ describes a system
(which could be large and complicated, such as a ``measuring apparatus")
at some initial time $t_i$, then, should a human calculator wish to
know the probability of a quantum event outcome $\ket{\phi(t_f)}$, he or she can calculate
$|\bra{\phi(t_f)} U(t_f,t_i) \ket{\psi(t_i)}|^2$, where $U(t_f,t_i)$ is the propagator.
However, it does not follow that $\ket{\psi(t)}$, at any given time $t$,
completely captures what
can be said about the system, because it does not in itself
contain the information that an event is taking place with $\ket{\phi(t_f)}$
in the final preferred basis. That is, the quantum system cannot
`know', from the information in $\ket{\psi(t)}$ at any given time $t$ alone, what
sort of non-unitary evolution it is participating in. Therefore to call
$\ket{\psi(t)}$ a ``state'' is a misnomer. 

Sufficient information (to allow the non-unitary evolution to be specified)
is contained in the unitary evolution of $\ket{\psi(t)}$
extended over time, through the topology. According to the quantum event
postulate, the non-unitary evolution is
the one actually undertaken by the system, but the set of paths extended over
spacetime (i.e. the unitary evolution) is what determines the
possible outcomes (in the strong sense of picking the actual preferred basis)
and their probabilities. Therefore the classical notion of an instantaneous
``state" has to be abandoned (it will emerge as a good approximation in
circumstances corresponding to classical-like evolution). The closest
equivalent to a ``state" at time $t \ge t_i$ is perhaps offered by the set of pairs:
\be
{\cal S} = \left\{ \left(
\ket{\phi_n(t)}, |\bra{\phi_n(t_f)} U(t_f,t_i) \ket{\psi(t_i)}|^2
\right), \; n=1,2,\ldots \right\},   \label{phiU}
\ee
where $\ket{\phi_n(t_f)},n=1,2,\ldots$ is the preferred basis selected by $U$, 
$\ket{\phi_n(t)} = U^{\dagger}(t_f,t) \ket{\phi_n(t_f)}$, and $t \le t_f$.
Each pair $(\ket{\phi},P)$ in the set consists of a ket and its probability.
 
In summary, an event outcome
(for example, $A$ and $B$ in $\ket{\downarrow\downarrow}$ or
$\ket{\uparrow\uparrow}$) is established both by the interactions between particles
and by their context, i.e. future interactions with other systems.
The unitary future says `these are the kinds of quantum event outcome
which may occur' (because their relative phases are going to be inconsequential
in any case), and the unitary evolution from the
past allows the probabilities to be obtained. This information is combined in the
non-unitary quantum event. This is similar to CI, but we have replaced the
appeal to `measurement' or `classical' systems by postulates 2 and 3:
an appeal to the atemporal character of unitary behaviour, and yet
an arrow of time revealed by the structure, and especially the topology,
of that behaviour. Atemporal behaviour was hinted at in classic
paradoxes such as the ``delayed choice",
but here we extend it right into the workings of the `classical' device.

Our interpretation therefore requires the idea of atemporality, and
this naturally leads to the consideration of non-locality. It is well-known that
the combination of objective reality and Einstein locality is compromised in
quantum theory: this is essentially what the Bell argument and related
experiments demonstrate. Most authors conclude that locality is compromised,
but some, notably Rovelli \cite{96Rovelli}
and perhaps Mermin \cite{98Mermin} propose the opposite conclusion.
It is certain that locality should not be lightly jettisoned. It is
deeply ingrained in physics, most notably in general relativity.

I share the general unease with the phrase `collapse of the wavefunction'.
In a CI-like interpretation, one can save locality by blurring objectivity
a little. For, during the unitary part of a quantum event, the relevant system
is described equally well by two kets: that evolved forwards from the
past, and that evolved backwards from the future. We already presented this
fact in equation (\ref{phiU}). In the example of
figure 1a, given the future context,
the initial condition of $A$ and $B$ could be one of
$\ket{\downarrow} \otimes \ket{\downarrow}$ or
$\ket{\uparrow} \otimes \ket{\downarrow}$, instead of
$\ket{\leftarrow} \otimes \ket{\downarrow}$ as we said before.
If $A$ and $B$ adopt one of these, their subsequent interaction
will put them in one of $\ket{\downarrow\downarrow}$ or $\ket{\uparrow\uparrow}$.
Therefore the probabilistic aspect of a quantum event does not have
to be located across a space-like interval in the final conditions, it
can equally be located at a point in the initial conditions. More
generally, because quantum events are extended not in an arbitrary way,
but along the worldlines, a local (but atemporal)
interpretation of the way correlations come about is always available.

The above has some elements in common with the ``consistent histories"
or ``decoherent histories" formulation or interpretation of quantum
mechanics. In common are the idea that the fundamental objects of
the theory are extended in time, and the assertion by postulate that
the theory associates physical reality with just those sets of possibilities
in which classical sum rules for probabilities are obeyed. However,
that condition is not very restrictive, and in particular it would not
on its own allow one to prefer one basis over another in 
equations (\ref{sep}), (\ref{ent}). The present treatment
makes (by postulate) a stronger statement, by preferring
separable outcomes over inseparable ones, where both are consistent
with the classical sum rules. This stronger condition might suffice
to settle the set-selection questions raised in \cite{96Dowker,95Dowker,98Kent}.
Discussions of consistent histories are sometimes ambiguous about
whether there exists a single evolving world with a definite past \cite{96Dowker,04Halliwell}.
In the present discussion it is assumed that there is a single evolving world
and by examining individual (extended) quantum events we can deduce what
aspects of the past are definite.  

\section{Microscopic mechanism to detect spacetime loops}  \label{s:trans}

One can adopt the ``no-loop" quantum event postulate simply as a statement
without putting forward a mechanism, but if one could find a mechanism
then it would clarify matters.
In this section I sketch a speculative account of how one might think
about the presence or absence of spacetime loops in the future being
probed by particles in the present. The sketch is inspired
by Cramer's transactional interpretation. Something like this is
needed to complete it.

\begin{figure}[ht!]
\centerline{\resizebox{!}{0.25\textwidth}{\includegraphics{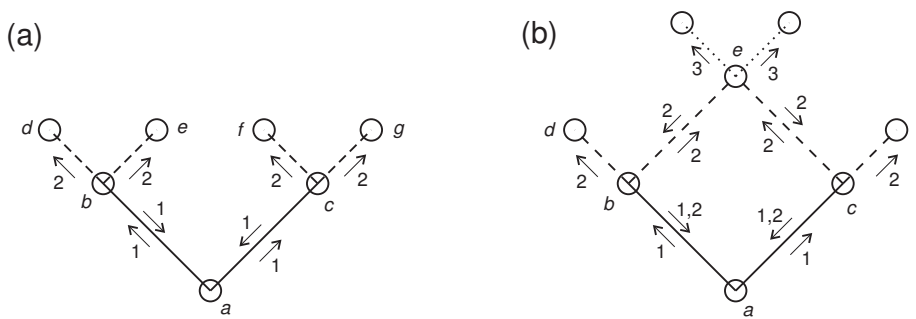}}}
\caption{Spacetime diagrams, showing a sketch of a
transactional-type mechanism for detecting
spacetime loops and thus deciding what type(s) of event can happen.
An emitter $a$ emits offer waves (OW) to
absorbers $b$ and $c$, which in turn emit second-order OWs. (a) No absorber
receives two 2nd-order OWs, so no 2nd-order CWs return to $b$ and $c$, and
these emit only 1st-order CWs to $a$. (b) An absorber $e$ receives two
2nd-order OWs. It generates a 3rd-order OW and 2nd order CWs to $b$ and $c$.
The 2nd-order OWs are not fully absorbed, so $b$ and $c$ pass both 1st and
2nd order CWs to $a$.
}
\end{figure}

Let us adopt Cramer's language of offer wave (OW, retarded)
and confirmation wave (CW, advanced), and
consider two simple scenarios in one dimension (figure 2). An emitter $a$
emits an
OW propagating left and right towards two absorbers $b$ and $c$.
In case
(a), the absorbers and subsequent OW/CW worldlines
and absorbers are so
placed that there is no spacetime loop (of a type relevant to
the event postulate). In (b) there is a spacetime loop.

We adopt the language of a sequence of events for purposes of discussion,
but of course we are referring to a single atemporal event. We will refer
to the initial offer wave emitted by $a$ as a `first order' OW.

Upon receiving a first-order OW, each absorber $b$, $c$ emits a
further OW (retarded, propagating
forwards in time), but of a higher-order (`second order')
type, describing entanglement between itself and other absorbers.
Second-order OWs can only be absorbed in groups of more than one.
In case (a) there is no absorber
for the higher-order wave, in case (b) there is. In case (a), upon receiving
no CW for the second-order OW, $b$ and $c$ proceed to emit
a first-order CW back towards the emitter $a$, and the transaction proceeds
as in Cramer's description. In case (b), one of the further absorbers, $e$,
receives two second-order OWs from $b$ and $c$. It emits a third-order
OW into the future, but receives no CW for that. Then it returns
a second-order CW to $b$ and $c$, which pass it on to $a$.
The 2nd-order transaction with $e$
is not guaranteed to happen: the conserved quantities (energy, momentum, etc.)
might be passed on to $d$ or $f$, so
$b$ and $c$ also return first-order CWs to $a$, of smaller amplitude than
in case (a).
$a$ then forms either a first-order transaction with one of $b$ and $c$, or
a second order transaction with $e$. The latter case involves an entanglement
between $b$ and $c$, of the form
\be
\left( \ket{g}_b\ket{e}_c + e^{i\phi} \ket{e}_b\ket{g}_c \right) / \sqrt{2}
\ee
where $\ket{g}$, $\ket{e}$ are ground and excited states.

To keep things simple, here I assumed there were no further
possible transactions in the future. If there were, then both cases
would be more involved: one would need to allow for
the possibility of entanglement involving all of $d,e,f,g$ for example,
and therefore these would all emit third-order OWs. I repeat that
this account is only a sketch and could be disregarded without impact
on the rest of the discussion.

\section{2nd Law of thermodynamics} \label{s:thermo}

The statement I have called the postulate of irreversibility is closely linked
to, though not identical with, the 2nd Law of thermodynamics. The latter
is a stronger statement, as is clear from the Carath\'eodory version:
{\em In the neighbourhood of any equilibrium state of a system
there are states which are inaccessible by an adiathermal process.} \cite{BkPippard}.
However, the present discussion suggests that the 2nd Law should be regarded
from a new perspective.

The 2nd Law of thermodynamics, and the concept of entropy, are commonly
regarded as useful devices rather than insights into the basic fabric of
the world. That is to say, the usual perspective is that the basic equations
of motion are reversible, but given that one would rather not keep track
of all the microscopic details of particle motion, it is useful to study
the average behaviour of large-scale quantities such as entropy. The concept
of entropy is then an idealization which applies in the thermodynamic limit, but
is not needed to describe the underlying motions of the constituent particles
of any real system.

However, according to the quantum event postulate, microscopic evolution
is a subtle mixture of reversible and irreversible components. The irreversible
part involves the system `forgetting' an alternative outcome, and the
relative phase which would otherwise
be retained. Therefore entropy is a basic ingredient in any
fundamental physical theory, and not merely a calculational device.

\section{Falsification, outlook} \label{s:false}

In this section, I will discuss ways in which the ideas presented here
could be falsified if they are in fact wrong, and I will outline avenues for
further investigation.

Interpretations of quantum theory which agree with the standard
predictions can't be distinguished by experimental tests
within the realm of applicability of the theory. However, they
can offer guidance in formulating generalizations or extensions
of known physics. The most obvious area of study in which this might be
relevant is in efforts to unify quantum theory and general relativity,
such as string theory, twistor theory and loop quantum gravity.

The contextuality principle may be regarded as a type of symmetry principle,
stating that any fundamental physical theory must not implicitly provide
more information about isolated systems than is allowed by the principle.
I believe that standard quantum theory is in agreement with it, and
indeed this appears to be a deep aspect of quantum theory that any
future theory will share.

The quantum event (no-loop) postulate might appear to be making a
circular argument. However, it is not.
It claims, roughly speaking, that the moon is in one place as long as no-one
will ever check to see if it is in two places. This is the strong statement.
The weak statement is, roughly, that the moon is in one place as long as no-one
could ever check to see if it is in two places. It also adds the
hypothesis that natural processes tend towards separable states
(other things being equal).

The combination of the ideas in this paper emphasizes spacial degrees of freedom,
since these are required for the definition of separability, and
only the results of interactions between spatially separable systems
determine the physical reality.
The interpretation thus places space, and spacetime, at a more profound level
in the description of things than Hilbert space. That is, the notions of Hilbert
space and operators should be regarded as useful tools, but the fundamental
equations are best formulated directly in terms of motion in spacetime, such
as for example in Feynman's
spacetime approach to quantum mechanics \cite{48Feynman}. If the more general
forms of quantum theory which (we hope) will be discovered in the future do
not have this feature then the ideas would be undermined.

On the other hand, positive evidence for the correctness of this approach
would be furnished if it and a quantum gravity theory were mutually supportive.
For example, it might clarify which aspects of
a highly complex theory represent identifiable elements of physical
reality, and which are part of the mathematical background.
If quantum effects in the structure of
spacetime provided the $A_{\rm max}$ invoked in the weak statement
of the quantum event (no-loop) postulate, that is, a limit to the
definition of phase in quantum interferences of large systems, then this
would lend (modest) support to the postulate. Quantization of spacetime
might also permit a natural way to identify, or place boundaries around,
the start and finish of quantum events.

Like consistent/decoherent histories,
Cramer's transactional description has in common with the ideas put
forward here that it emphasizes
spacetime above Hilbert space, c.f. \cite{96Dowker,97Sorkin}.
It is, I have argued, in need of
completion, but it is valuable because it is
mathematically precise, and the sketched suggestion for completing it
given in section \ref{s:trans} does not do justice to that. If this sketch
could be replaced by a thorough analysis, it would provide
one way to make a more detailed statement of the no-loop postulate.
Conversely, if this were not possible, it would tend to undermine the postulate.

If there is a further dynamical collapse mechanism to be discovered,
the present discussion suggests that it would involve or promote a
resistance to non-separable physical configurations, and would have
an atemporal element.

The discussion drew on the concepts both of entanglement
and of irreversibility. This suggests that there may exist
a unifying theoretical structure which brings together these two concepts more
thoroughly. For example, one might link the maximization of separability of
a given pair of systems with the maximization of entropy in the rest of
the universe. A weakness of the discussion provided here
is the lack of information on how
one is to calculate separability in cases more complicated than the simple
example treated. This is a hard problem and is currently undergoing
extensive study in the quantum information community \cite{BkNielsen}.

\section{Conclusion}  \label{s:conc}

The combination of ideas in this paper may be called a
`contextual, temporal' interpretation of quantum mechanics.
According to our treatment,
{\em it is possible to have objective physical events
if and only if some processes are non-reversed.}
The tension between reversibility and irreversibility is the same as
the tension between contextuality and objectivity.

In summary, the principle of contextuality states that isolated entities cannot
have well-defined properties in and of themselves, and a
basic theory ought not to imply that they can.
It provides a symmetry principle which, if correct,
will be respected by basic theories in physics.
Quantum mechanics is a formalism in which this principle
finds mathematical expression through the concept of entanglement.
Because interactions and correlations are more fundamental than the
entities interacting and correlated,
physical entities have to be considered in groups of at least three in order
to allow statements about what transpires: two to have an interaction,
and a third to be influenced by the result.

When placed in the context of the actual evolution of the universe,
systems can acquire properties through a symmetry-breaking process.
Physical reality constitutes a sequence of random
non-unitary evolutions between physical
configurations. The configurations and their probabilities are determined
by a mathematical apparatus describing quantum amplitudes
and unitary evolution along all paths.
The quantum amplitudes contain more information than finds
physical expression, however: when the paths form closed loops
(in the sense of no `which-path' information), the relative phases
of the amplitudes can influence events, when they do not, the relative phases
cannot. The physical behaviour is so constituted as to express
this in as economic a way as possible. That is, the random quantum events
evolve systems to new configurations drawn from a set whose quantum amplitudes
have relative phase that will never be physically relevant.
This is possible because non-reversed processes, that is, paths
which never form closed loops, occur.

I thank Harvey Brown for commenting on an early
version of the manuscript, Fay Dowker for helpful comments and editing,
and Ferdinand Schmidt-Kaler for arranging
an extended working visit during which these ideas matured.
This work was supported by the Conquest network of the EC RTN programme,
and the EPSRC through QIP IRC.

\bibliographystyle{unsrt}
\bibliography{myrefs}

\end{document}